%% file: Popularity_Mobile_User_Real_Data_2019_04_14.tex
\documentclass[conference]{IEEEtran}

\usepackage{amsmath}
\usepackage{amsfonts}
\usepackage{amssymb}
\usepackage{graphicx}
\usepackage{epstopdf}
\usepackage{caption}
\usepackage{subcaption}
\usepackage{cite}
\usepackage{algorithmicx}
\usepackage[ruled]{algorithm}
\usepackage{algpseudocode}
\usepackage{blindtext}
\usepackage{enumitem}
\usepackage{pdfpages}
\usepackage{bbm}
\usepackage{amsthm}
\usepackage{cite}
\usepackage{url}
\usepackage{tabularx}
\usepackage{bibentry}

\usepackage{changes}
\definechangesauthor[color=red,name={Mingyue Ji}]{MJ}

\input{macros}

\begin{document}

\title{Performance of Caching-Based D2D Video Distribution with Measured Popularity Distributions}

\author{Ming-Chun Lee, Mingyue Ji, Andreas F. Molisch, and Nishanth Sastry
}

\maketitle

\newtheorem{theorem}{\bf{Theorem}}

\newtheorem{lemma}{\bf{Lemma}}
\newtheorem{prop}{\bf{Proposition}}
\newtheorem{definition}{\bf{Definition}}
\newtheorem{example}{\bf{Example}}

\begin{abstract}
On-demand video accounts for the majority of wireless data traffic. Video distribution schemes based on caching combined with device-to-device (D2D) communications promise order-of-magnitude greater spectral efficiency for video delivery, but hinge on the principle of ``concentrated demand distributions.'' This paper presents, for the first time, the analysis and evaluations of the throughput--outage tradeoff of such schemes based on {\em measured cellular demand distributions}. In particular, we use a dataset with more than 100 million requests from the BBC iPlayer, a popular video streaming service in the U.K., as the foundation of the analysis and evaluations. We present an achievable scaling law based on the practical popularity distribution, and show that such scaling law is identical to those reported in the literature. We find that also for the numerical evaluations based on a realistic setup, order-of-magnitude improvements can be achieved. Our results indicate that the benefits promised by the caching-based D2D in the literature could be retained for cellular networks in practice.\let\thefootnote\relax\footnote{M.-C. Lee and A. F. Molisch are with Department of Electrical and Computer Engineering, University of Southern, Los Angeles, CA 90089, USA (email: mingchul@usc.edu, molisch@usc.edu).}\addtocounter{footnote}{-1}\let\thefootnote\svthefootnote
\let\thefootnote\relax\footnote{M. Ji is with Department of Electrical and Computer Engineering, University of Utah, Salt Lake City, UT 84112, USA (email: mingyue.ji@utah.edu).}\addtocounter{footnote}{-1}\let\thefootnote\svthefootnote
\let\thefootnote\relax\footnote{N. Sastry is at the Department of Informatics, King's College London, London, UK (e-mail: nishanth.sastry@kcl.ac.uk).}\addtocounter{footnote}{-1}\let\thefootnote\svthefootnote
\let\thefootnote\relax\footnote{This work was supported in part by the National Science Foundation (NSF) under CNS-1816699 and CCF-1423140.}\addtocounter{footnote}{-1}\let\thefootnote\svthefootnote 
\end{abstract}

\section{Introduction}

Wireless data traffic is anticipated to increase rapidly per year in the near future. The on-demand video traffic, which accounts for about $2/3$ of all wireless data \cite{Cisco:5G}, is the primary reason for such growth, and video service has become the ``killer application'' for both 4G and 5G cellular systems. Finding cost-effective approaches to distribute videos in cellular networks is thus one of the most important and urgent tasks for the development of mobile systems.

Traditional methods for video distribution transfer files and provide on-demand streaming services via conventional unicast, meaning that they do not distinguish video traffic from any other traffic. Consequently, they rely on the general throughput enhancement methods of cellular networks such as network densification, HetNets \cite{Andrews_2013}, massive MIMO, and use of additional spectrum (in particular mm-wave bands \cite{Rappaport_2013}). However, these approaches tend to be either very expensive, and/or not scalable.

Compared to other data traffic, on-demand video has two unique properties: (i) high concentration of the popularity distribution, i.e., a small number of videos accounts for the majority of the video traffic, and (ii) {\em asynchronous content reuse}, i.e., video files are watched by different people at different times.\footnote{The latter property distinguishes 
video streaming services such as Netflix, Amazon Prime, Hulu, and Youtube, from the traditional broadcast TV, which achieved high spectral efficiency by forcing viewers to watch particular  videos at prescribed times.} This enables the networks to {\em convert memory into bandwidth} \cite{Ji:Th_Out_toff} by employing caching as part of the video distribution process. Such an approach is appealing because bandwidth is limited and expensive, while memory is a cheap and rapidly growing hardware resource. Commonly used caching approaches include selfish on-device caching \cite{Gol:Dcache1}, femtocaching \cite{Gol:femtocaching}, coded mutlicasting \cite{Maddah-Ali:CCache,Ji: Fund_D2D}, and caching combining file transfer with device-to-device (D2D) \cite{Gol:Dcache1,Gol:femtocaching,Ji:Th_Out_toff,Ji:Dcache}. Since the last method, namely the caching-based D2D, can provide not only appealing scaling laws (network throughput increasing linearly with the number of devices) but also robustness in realistic propagation conditions \cite{Ji:Dcache}, this paper will therefore place the focus on it.   

The fundamental implementation of caching-based D2D is as follows. Each device caches a subset of the popular video files based on a caching policy.\footnote{A caching policy could be either deterministic or random, and is generally a function of the video popularity distribution and other system parameters.} When a user requests a file, it might either already be in this user's cache (in which case it is retrieved from there), or is obtained from a near-by device through spectrally efficient, short-distance D2D communications. This approach was first suggested by one of the authors in \cite{Gol:Asilomar_2011, Gol:femtocaching}. The information-theoretic scaling laws was later developed by two of us in \cite{Ji:Th_Out_toff, Ji:Dcache,Ji: Fund_D2D}, showing the significant benefits; various theoretical and practical aspects have been studied, e.g., in \cite{Chen:Dcache,Chen:D2D_Coop,Guo:Throughput,Lee:Indi_pre_model_ToN}. 

Most existing papers are based on the assumption that the popularity distribution has the shape of the Zipf distribution (essentially a power law distribution), with a Zipf parameter typically between $0.6$ and $2$. However, this assumption was based on observations in a {\em wired} network \cite{UMass} with Youtube videos. A recent investigation \cite{Notre-Dame} into wireless popularity distributions of general content showed little content reuse. It is noteworthy that - as the authors of the paper point out - this investigation could not identify video reuse, since video connections were run via a secure https connection, so that the content of the videos could not be determined.\footnote{The paper has been sometimes {\em misinterpreted} as indicating that there is little video reuse.} The question thus remains open whether caching-based D2D video distribution can provide cellular networks in practice the benefits that the theory promises.

To answer the question, we first use an extensive dataset of the BBC iPlayer, one of the most popular video distribution service in the UK, to find the {\em measured} video popularity distribution of cellular users. Through appropriate processing, we extract the popularity distribution for the videos watched only via cellular connections (these might be different from the files watched through wired/WiFi connections). We find that, rather than using a Zipf distribution, a Mandelbrot-Zipf distribution (MZipf) \cite{Hefeeda:P2P} with a somewhat lower concentration provides better description for the extracted popularity distribution. Based on the MZipf distribution, we then analyze the throughput--outage scaling law and numerically evaluate the performance of the caching-based D2D network. Our results show that, despite the lower concentration of the MZipf popularity distribution, the throughput--outage scaling law can be identical to the one provided for the Zipf popularity distribution \cite{Ji:Th_Out_toff} when the MZipf popularity distribution is sufficiently skewed.\footnote{The condition that a MZipf popularity distribution is sufficiently skewed is not difficult to realize in practice as we will see from the results later} Besides, numerical experiments show that caching-based D2D schemes can provide orders of magnitude improvement of throughput for a given outage probability. Both the theoretical and numerical results indicate that the benefits of the cache-based D2D can be retained for mobile video distribution considering a more practical popularity distribution.

The remainder of the paper is organized as follows: Sec. II presents the dataset for video demands and the extracted popularity distribution as well as the modeling. We present the throughput--outage analysis results in Sec. III. Sec. IV summarizes the simulation results based on the measured popularity distribution. Sec. V concludes this paper.  

\section{The Measured Data and Modeling Results}

This paper uses an extensive set of real-world data, namely the dataset of the BBC iPlayer \cite{Karamshuk:PreCache,Lee:Indi_pre_model_ToN}, to obtain realistic video demand distributions. The BBC iPlayer is a video streaming service provided by BBC (British Broadcasting Corporation) that provides video content for a number of BBC channels without charge. Content on the iPlayer is basically available for up to 30 days depending on the policies. We consider the dataset covering July 2014, which includes $190,500,463$ recorded access sessions. In each record, access information of the video content contains two important columns: \textit{user id} and \textit{content id}. \textit{user id} is based on the long-term cookies that uniquely (in an anonymized way) identify users. \textit{content id} is the specific identity that uniquely identifies each video content separately. Although there are certain exceptions, \textit{user id} and \textit{content id} can generally help identify the user and the video content of each access. More detailed descriptions of the BBC iPlayer dataset are in \cite{Karamshuk:PreCache,Lee:Indi_pre_model_ToN}.

To facilitate the investigation, preprocessing is conducted on the dataset. By observation, we notice that a user could access the same file multiple times, possibly due to temporary disconnnections from Internet and/or due to temporary pauses raised by users when moving between locations. Since a user is generally unlikely to access the same video after finishing to watch the video within the period of a month \cite{Lee:Indi_pre_model_ToN}, we consider multiple accesses made by the same user to the same file as a single unique access. 

We then separate the data required by cellular users from those requested via cabled connections or personal WiFi, resulting in $689,461$ different unique accesses (requests) among $327,721$ different users. We do this by observing the Internet gateway through which the requests are routed. For example, one of the major cellular operators in the UK has several gateways across the country and all cellular request go through these gateways. This on one hand allows an easy separation, but on the other hand does not allow a precise localization of the requests: as outlined below, we can only divide the whole country into 3 (unequal) regions, and two of them might have approximately the same size as a city. We will thus make in the following the assumption that the popularity distribution at each location follows the global (over a particular region) popularity distribution.\footnote{We understand that our results might not fully represent the results of a small area, e.g., a cell. However, those are the best indication currently available, because, to the best of our knowledge, there does not exist publicly available data for {\em video reuse} of {\em mobile} data {\em on a per-cell basis}.}

\begin{figure}
\centering
\includegraphics[width=7.0cm,keepaspectratio]{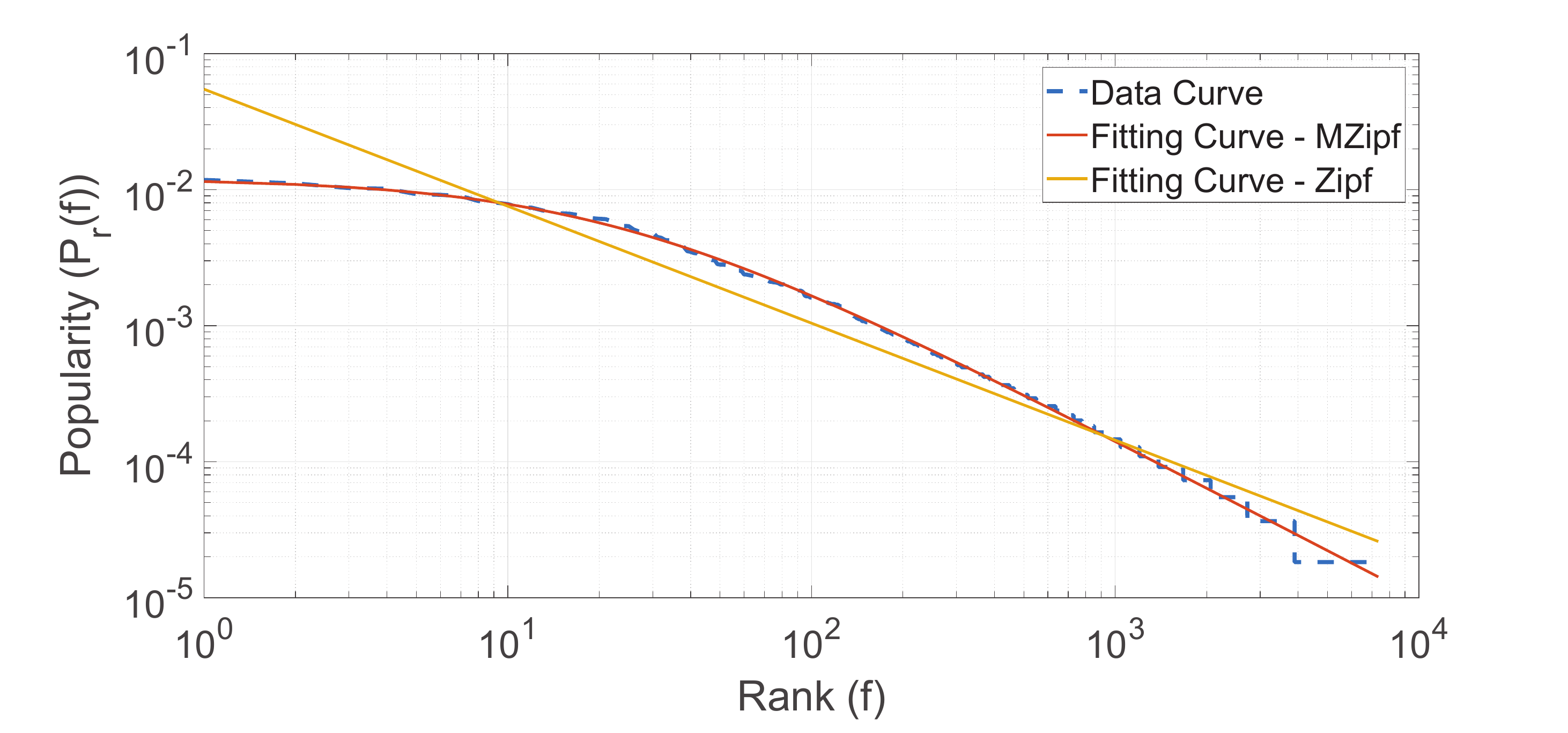}
\caption{Measured ordered popularity distribution of video files of the BBC iPlayer requested via one of the major cellular operator in July of 2014 in region 2. }
\label{fig:global_distribution2}
\vspace{-10pt}
\end{figure}
\begin{figure}
\centering
\includegraphics[width=7.0cm,keepaspectratio]{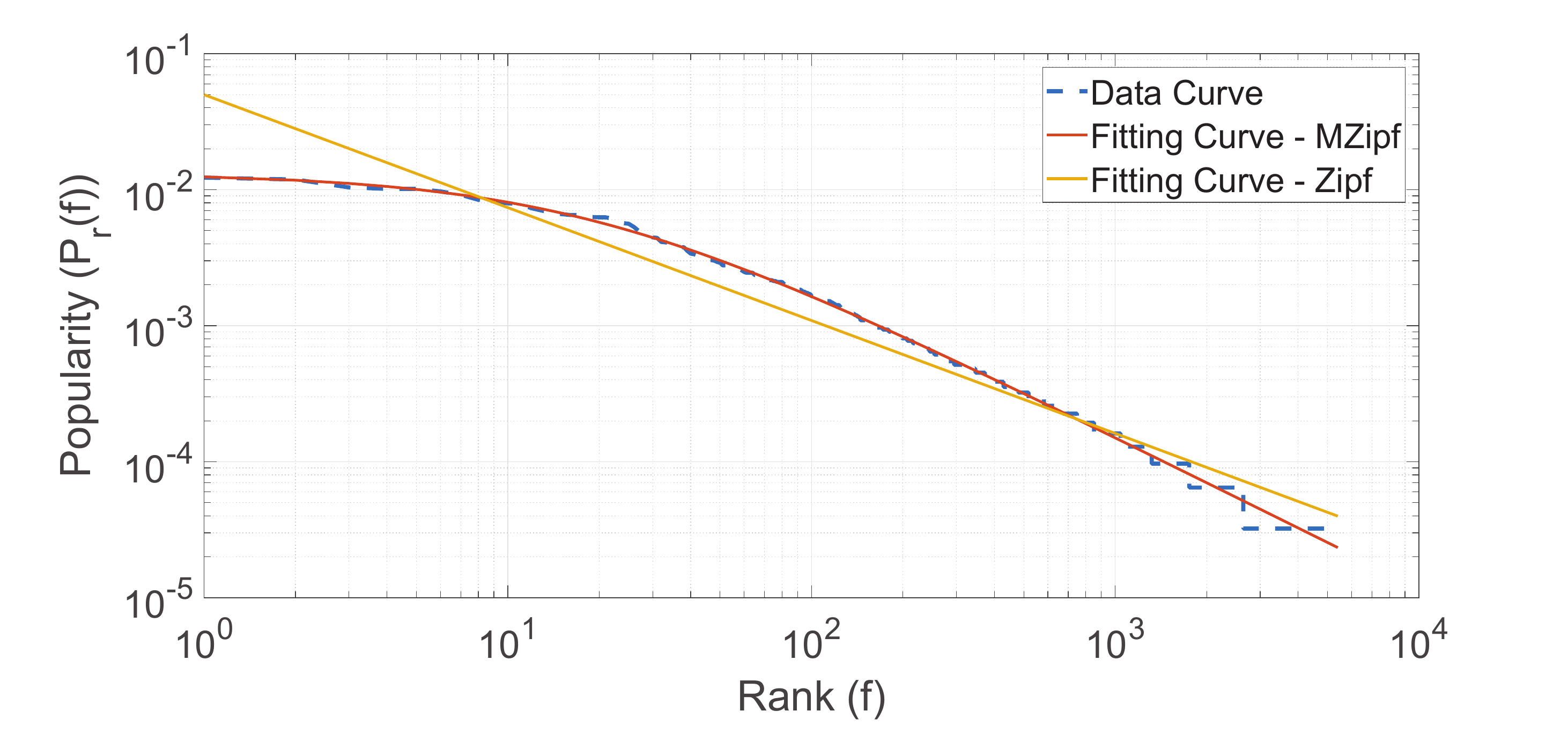}
\caption{Measured ordered popularity distribution of video files of the BBC iPlayer requested via one of the major cellular operator in July of 2014 in region 3. }
\label{fig:global_distribution3}
\vspace{-10pt}
\end{figure} 
Based on these data, we plot the global popularity distribution and find that the traditional Zipf model is not a good fit, see Figs. \ref{fig:global_distribution2} and \ref{fig:global_distribution3}. Rather, a MZipf distribution \cite{Hefeeda:P2P} provides a good approximation:
\begin{equation}
\label{eq:gener_pop}
P_r(f)=\frac{(f+q)^{-\gamma}}{\sum_{j=1}^{M} (j+q)^{-\gamma}},
\end{equation}
where $\gamma$ is the Zipf factor, and $q$ is the plateau factor. We note that the MZipf distribution degenerates to a Zipf distribution when $q=0$. 

A fitting that minimizes the Kullback-Leibler (KL) distance provides the following values:\footnote{The KL distance of a parameter set $\mathbf{x}$ is defined as $D_{\text{KL}}(\mathbf{x})=\sum_m p_m^{\text{data}}\log\frac{p_m^{\text{data}}}{p_m^{\text{model}}(\mathbf{x})}$.} in region 1, $q=34$, $\gamma=1.28$, and $M=18553$ (these values are very similar to the case considering whole of the UK), while region 2 has $q=22$, $\gamma=1.16$, and $M=7345$ and region 3 has $q=18$, $\gamma=1.11$, and $M=5405$. This implies that up to a breakpoint of approximately 20-30 files, the popularity distribution is approximately flat, and decays from there. Also importantly, we find that for less-popular files, there is an power law decrease with a $\gamma >1$. Since regions 2 and 3 cover smaller regions, and thus are expected to describe better the effects that might be encountered within a particular cell (though they are still much larger than a cell), we will use their corresponding parameters henceforth. 

\section{Achievable Throughput--Outage Tradeoff}
From the measured data, we understand that the MZipf distribution is more suitable for mobile data traffic. In this section, we thus provide the achievable throughput--outage tradeoff analysis considering the MZipf distribution.
\subsection{Network Setup}
In this section, we describe the network model used for the analysis and define the throughput--outage tradeoff, which is similar to the model used in \cite{Ji:Th_Out_toff}. Denote the number of users in the network as $N$. We assume a network where user devices can communicate with each other through direct links. We consider the transmission policy using {\em clustering}, in which the devices are grouped geographically into clusters such that any device within one cluster can communicate with any other devices in the same cluster with a constant rate $C$ bits/second/Hz, but not with devices in a different cluster. The network is split into equal-sized clusters. We adopt a grid network in which the users are placed on a regular grid. As a result, $g_c(M)\leq N\in\mathbb{N}$ determined as a function of $M$ and denoted as the cluster size in this paper, is the number of users in a cluster and is a parameter to be chosen in order to analyze the throughput--outage tradeoff.

Only one D2D link is allowed within an active cluster. Adjacent clusters are assumed to use different time/frequency resources. As a matter of fact, a ``reuse'' factor is used to combat inter-cluster interference. A spatial reuse scheme with Time Division Multiple Access (TDMA) is thus adopted.\footnote{We use TDMA only as convenient example. Any scheme that allocates orthogonal resources to clusters with different colors is aligned with our model.} Denoting $K$ as the reuse factor, such a reuse scheme evenly applies $K$ colors to the clusters, and only the clusters with the same color can be activated on the same time-frequency resource for D2D communications. A potential link in the cluster is said to exist as there is a user can find its desired file in the cluster through D2D communications. We then say that a cluster is {\em good} if it contains at least one potential link. Since there could exist multiple potential links in a cluster, we schedule the potential links of the same cluster with equal probability (or, equivalently, in round robin). Therefore all users have the same average throughput.

Although the assumptions above are made for the subsequent theoretical analysis, we actually can realize most of them in practice. Specifically, the adjustable size of the cluster can be implemented by adapting the transmit power - in other words, the transmit power is chosen such that communication between opposite corners of a cluster is possible. The link rate for the D2D communication is fixed when no adaptive modulation and coding is used, and of course this rate has to be smaller than the capacity for the longest-distance communication envisioned in this system. The signal-to-noise ratio (SNR) is determined by the pathloss; small-scale fading can be neglected since in highly frequency-selective channels, the effects of this fading can be eliminated by exploiting the frequency diversity. 

It must be emphasized that the above network is not optimum for D2D communications. Suitable power control, adaptive modulation and coding, etc., could all increase the spectral efficiency. However, our model provides both a useful lower bound on the performance as well as analytical tractability, which is important for comparability between different schemes. The information theoretical optimal throughput-outage tradeoff analysis is beyond the scope of this paper. To provide support for our theory, we will later provide numerical simulations  considering a practical setup in Sec. VI.

In this section, our goal is to provide the asymptotic analysis when $N\to\infty$, $M\to\infty$, and $q\to\infty$.\footnote{Based on the experimental results, $\gamma$ changes within a (small) finite range, i.e., does not go to infinity, as $M$ increases. We therefore approximate $\gamma$ as a fixed constant for the sake of analysis.} We denote $S$ as the cache memory in a user device, i.e., a user can cache up to $S$ files.\footnote{We consider $S$ as a fixed network parameter not to grow to infinity as $N\to\infty$, $M\to\infty$, and $q\to\infty$.} The aggregate memory in a cluster is thus $Sg_c(M)$. An independent random caching policy is adopted for users to cache files. Denote $P_c(f)$ as the probability of caching file $f$, where $0\leq P_c(f)\leq 1$ and $\sum_{f=1}^M P_c(f)=1$. Using such caching policy, each user caches each file independently at random according to $P_c(f)$.\footnote{A user might cache the same file multiple times under this caching policy, and this policy is used for the sake of analysis.}

Given the popularity distribution $P_r(\cdot)$, caching policy $P_c(\cdot)$, and transmission policy, we define the average throughput of a user $u$ as $\overline{T}_u=\mathbb{E}\left[ T_{u} \right]$, where $T_{u}$ is a throughput realization of user $u$, and the expectation is taken over the realizations of the cached files and requests.  The minimum average throughput is $\overline{T}_{\text{min}}=\displaystyle{\min_{u}}\overline{T}_u=\overline{T}_u$ due to the symmetry of the network (e.g., round robin scheduling). We define the number of users in outage $N_o$ as the number of users that cannot find their requested files. Thus the average outage is:
\begin{equation}\label{eq:Def_Outage}
p_o=\frac{1}{N}\mathbb{E}\left[N_o\right]=\frac{1}{N}\sum_{u}\mathbb{P}\left( \overline{T}_u=0\right)=1-P_u^c,
\end{equation}
where $P_u^c$ is the probability that a user $u$ can find its desired file in a cluster. Due to the symmetry of the network, $P_u^c$ is the same for all users. $P_u^c$ is also called ``hit-rate'' in some literature \cite{Chen:Dcache}. We note that our network setup for the theoretical results in this section follows the framework in \cite{Ji:Th_Out_toff}. Thus please refer to \cite{Ji:Th_Out_toff} for more rigorous descriptions. Following above definition and setup, we shall observe that both $\overline{T}_{\text{min}}$ and $p_o$ generally decreases with respect to $g_c(M)$, but we want $\overline{T}_{\text{min}}$ to be large while $p_o$ to be small. We thus aim to characterize the achievable throughput--outage tradeoff:

{\em Definition \cite{Ji:Th_Out_toff}:} For a given network and popularity distribution, a throughput--outage pair $(T, P_o)$ is achievable if there exists a caching policy and a transmission policy with outage probability
$p_o\leq P_o$ and minimum per-user average throughput $\overline{T}_{\text{min}}\geq T$.

\subsection{Throughput--Outage Tradeoff Results}
We now present the achievable throughput--outage tradeoff of caching-based D2D considering MZipf popularity distribution.\footnote{Scaling law order notation: given two functions $f$ and $g$, we say that: (1) $f(n)=\mathcal{O}(g(n))$ if there exists a positive constant $c$ and integer $N$ such that $\vert f(n)\vert\leq cg(n)$ for $n>N$. (2) $f(n)=o(g(n))$ if $\lim_{n\to\infty}\frac{f(n)}{g(n)}=0$. (3) $f(n)=\Omega(g(n))$ if $g(n)=\mathcal{O}(f(n))$. (4) $f(n)=\omega(g(n))$ if $g(n)=o(f(n))$. (5) $f(n)=\Theta(g(n))$ if $f(n)=\mathcal{O}(g(n))$ and $g(n)=\mathcal{O}(f(n))$.} In this paper, we only present the results with $\gamma>1$, which is aligned with the measurement results.\footnote{We consider $q=\mathcal{O}(M)$ because, by definition, the MZipf distribution would converge to simple uniform distribution when $q=\omega(M)$. Besides, as a matter of practice, we can see from results in Sec. II that $q$ is much smaller than $M$. Note that we view the case that $q=\Theta(1)$ is a constant simply as a degenerate case of our results.} The comprehensive investigation of the achievable throughput--outage tradeoff is provide in \cite{Lee:throuOutage}, which covers generally all regimes we are interested in. Besides, due to page limitation, we omit all the proofs of the corollaries and theorems in this paper; their proofs are relegated to \cite{Lee:throuOutage}.

Under the network setup considered in Sec. III.A, we determine the throughput--outage tradeoff by adopting the caching policy maximizing $P_u^c$ and by adjusting the cluster size $g_c(M)$. We first provide the following theorem to describe the optimal caching policy:

{\em Theorem 1:} We define $c_2=qa'$, where $a'=\frac{{\gamma}}{S(g_c(M)-1)-1}$, and $c_1\geq 1$ is the solution of the equality $c_1=1+c_2\log\left(1+\frac{c_1}{c_2}\right)$. Let $M\to\infty$, $N\to\infty$, and $q\to\infty$. Suppose $g_c(M)\to\infty$ as $M\to\infty$, and denote $m^*$ as the smallest index such that $P_c^*(m^*+1)=0$. Under the network model in Sec. III.A, the caching distribution $P_c^*(\cdot)$ that maximizes $P_u^c$ is:
\begin{equation}
P_c^*(f)=\left[1-\frac{\nu}{z_f} \right]^+,f=1,...,M,
\end{equation}
where $\nu=\frac{m^*-1}{\sum_{f=1}^{m^*}\frac{1}{z_f}}$, $z_f=\left(P_r(f)\right)^{\frac{1}{S(g_c(M)-1)-1}}$, $[x]^+=\max(x,0)$, and
\begin{equation}
m^*=\Theta\left(\min\left(\frac{c_1Sg_c(M)}{\gamma},M \right)\right).
\end{equation}
\begin{proof}
See the Proof of Theorem 1 in \cite{Lee:throuOutage}.
\end{proof}

Observe that $P_c^*(f)$ is monotonically decreasing and $m^*$ determines the number of files whose $P_c^*(f)>0$. Besides, we can observe that $c_1\geq 1$ and $c_1=1$ only if $c_2=o(1)$. Furthermore, we can see that $c_1=\Theta(c_2)$ when $c_2=\Omega(1)$. Thus, when considering $q=\Omega\left(\frac{Sg_c(M)}{\gamma}\right)$ and $\frac{c_1Sg_c(M)}{\gamma}<M$, we obtain $m^*=\Theta(\frac{c_1Sg_c(M)}{\gamma})=\Theta(\frac{c_2Sg_c(M)}{\gamma})=\Theta(q)$. Combining above results, Theorem 1 indicates that the caching policy should cover at least up to the file at rank $q$ (order-wise) in the library. This is intuitive because the MZipf distribution has a relatively flat head and $q$ characterizes the breaking point.

Using Theorem 1, we then characterize $P_u^c$, i.e., the probability that a user can find the desired file in a cluster, in Corollaries 1 and 2:

{\em Corollary 1:} Let $M\to\infty$, $N\to\infty$, and $q\to\infty$. Suppose $g_c(M)\to\infty$ as $M\to\infty$. Consider $q=\mathcal{O}\left(\frac{Sg_c(M)}{\gamma}\right)$ and $g_c(M)<\frac{\gamma M}{c_1 S}$. Under the network model in Sec. III.A and the caching policy in Theorem 1, $P_u^c$ is given by (\ref{eq:Corollay_1}) on the top of next page.
\begin{figure*}
\begin{equation}\label{eq:Corollay_1}
P_u^c=\frac{ \left(\frac{c_1Sg_c(M)}{\gamma}+q\right)^{1-\gamma} }{(M+q)^{1-\gamma}-(q+1)^{1-\gamma}}-\frac{(1-\gamma)\left(\frac{c_1Sg_c(M)}{\gamma}+q\right)^{-\gamma}\left(\frac{c_1Sg_c(M)}{\gamma}\right)}{(M+q)^{1-\gamma}-(q+1)^{1-\gamma}}-\frac{(q+1)^{1-\gamma}}{(M+q)^{1-\gamma}-(q+1)^{1-\gamma}}.
\end{equation}
\end{figure*}

{\em Corollary 2:} Let $M\to\infty$, $N\to\infty$, and $q\to\infty$. Suppose $g_c(M)\to\infty$ as $M\to\infty$. Consider $q=\mathcal{O}\left(\frac{Sg_c(M)}{\gamma}\right)$ and $g_c(M)=\frac{\rho M}{c_1S}$, where $\rho\geq \gamma$. Define $D=\frac{q}{M}$. Under the network model in Sec. III.A and the caching policy in Theorem 1, $P_u^c$ is lower bounded by (\ref{eq:Corollay_2}) on the top of next page.
\begin{figure*}
\begin{equation}\label{eq:Corollay_2}
P_u^c\geq 1- \frac{(1-\gamma)e^{-(\rho/c_1-\gamma)}}{(1+D)^{1-\gamma}-(D)^{1-\gamma}}\left[(1+D)^{\frac{\gamma}{S(g_c(M)-1)-1}+1}-\left(D\right)^{\frac{\gamma}{S(g_c(M)-1)-1}+1}\right]^{-(S(g_c(M)-1)-1)}.
\end{equation}
\end{figure*}
\begin{proof}
See the proofs of Corollaries 1 and 2 in \cite{Lee:throuOutage}.
\end{proof}

By using Corollaries 1 and 2, we can then derive the following Theorem:

{\em Theorem 2:} Let $M\to\infty$, $N\to\infty$, and $q\to\infty$. Suppose $g_c(M)\to\infty$ as $M\to\infty$. Consider $M=\mathcal{O}(N)$, $q=\mathcal{O}\left(\frac{Sg_c(M)}{\gamma}\right)$, and $\gamma>1$. Under the network model in Sec. III.A, the throughput--outage tradeoff achievable by adopting the caching policy in Theorem 1 is characterized by two regimes:

\begin{enumerate}[label=(\roman*)]
\item When $g_c(M)<\frac{\gamma M}{c_1 S}$ and $q=\mathcal{O}\left(\frac{Sg_c(M)}{\gamma}\right)$. Define $c_6=\frac{q}{g_c(M)}$. The achievable throughput--outage tradeoff is
\begin{equation}
T(P_o)=\frac{C}{K}\frac{1}{g_c(M)}+o\left(\frac{1}{g_c(M)}\right),
\end{equation}
where $P_o=(c_6)^{\gamma-1}\frac{Sc_1+c_6}{\left(\frac{Sc_1}{\gamma}+c_6\right)^{\gamma}}$.
\item Define $D=\frac{q}{M}$. When $g_c(M)=\frac{\rho M}{c_1S}$, where $\rho\geq \gamma$, the achievable throughput-outage tradeoff is
\begin{equation}
T(P_o)=\frac{C}{K}\frac{Sc_1}{\rho M}+o\left(\frac{1}{M}\right),
\end{equation}
where $P_o=1-P_{u,\text{Cor2}}^c$ and $P_{u,\text{Cor2}}^c=(\ref{eq:Corollay_2})$.
\end{enumerate}
\begin{proof}
Regime 1 follows the same proof as Theorem 4 of \cite{Lee:throuOutage}, and regime 2 follows the same proof as for regime 3 in Theorem 2 of \cite{Lee:throuOutage}. Note that the proof for regime 3 in Theorem 2 of \cite{Lee:throuOutage} is feasible for both $\gamma>1$ and $\gamma<1$.
\end{proof}

When $q=o\left(\frac{Sg_c(M)}{\gamma}\right)$, Theorem 2 directly leads to Corollary 3 as following:

{\em Corollary 3:} Let $M\to\infty$, $N\to\infty$, and $q\to\infty$. Suppose $g_c(M)\to\infty$ as $M\to\infty$. Consider $\gamma>1$ and $q=o\left(\frac{Sg_c(M)}{\gamma}\right)$. Under the network model in Sec. III.A and the caching policy in Theorem 1, the achievable throughput--outage tradeoff is
\begin{equation}
T(P_o)=\frac{C}{K}\frac{1}{g_c(M)}+o\left(\frac{1}{g_c(M)}\right),
\end{equation}
where $P_o=o(1)$.

From Theorem 2 and Corollary 3, we observe that when $\gamma>1$ and the order of $q$ is no larger than the the order of the aggregate memory,\footnote{This condition is generally true since it requires only $q=\mathcal{O}(M)$, and we are not interested in the case that $q=\omega(M)$ as indicated in footnote 9.} we obtain a scaling law that is at least as good as $\Theta(\frac{S}{M})$ - the throughput scales inversely to the library size $M$ and scales linearly with respect to the memory size $S$. Such scaling law is identical to the scaling law when the Zipf popularity distribution is considered, and is also identical to the scaling laws of the coded multicasting \cite{Maddah-Ali:CCache} and harmonic broadcasting \cite{Juhn:HarmBroad}.\footnote{Please refer to \cite{Ji:Th_Out_toff} for detailed discussions.} Furthermore, when $q=o(M)$, the results suggest that we obtain a scaling law that is better than $\Theta(\frac{S}{M})$ but worse than $\Theta(\frac{S}{q})$. This better result is even significant in practice since we see from Sec. II that $q$ is generally much smaller than $M$. In summary, the above results indicate that the benefits of the caching-based D2D network considering a Zipf popularity distribution can be retained when the popularity distribution follows the more practical MZipf distribution.

\subsection{Finite-Dimensional Simulations}
\begin{figure}
\centering
\includegraphics[width=7.6cm,keepaspectratio]{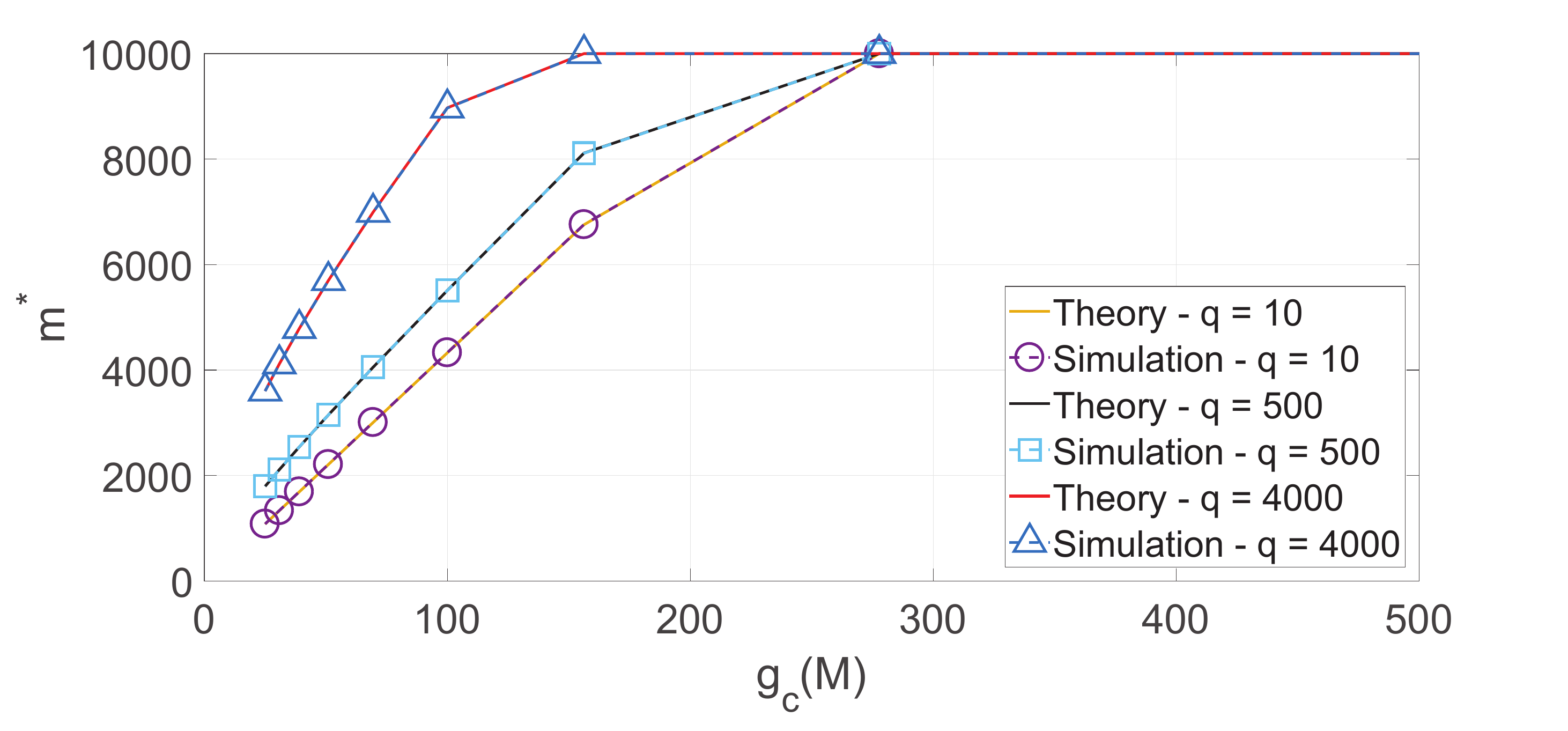}
\caption{Comparison between the theoretical $m^*$ and the simulated $m^*$ in a network with $\gamma=1.16$ and $M=10000$. }
\label{fig:M_star_Val}
\vspace{-15pt}
\end{figure}

Here we provide finite-dimensional simulations to compare theoretical and simulated results. In Fig. \ref{fig:M_star_Val}, we consider $\gamma=1.16$ and $M=10000$, and validate Theorem 1. Since the most critical part in Theorem 1 is the expression of $m^*$, we thus compare the theoretical $m^*$ in Theorem 1 with the $m^*$ obtained by numerically solving the Karush-Kuhn-Tucker (KKT) conditions. We observe that the theoretical results perfectly matches the simulated results even with finite dimensional setups. Simulations to compare between the theoretical and simulated throughput--outage tradeoff are also provided. Though the results are not shown in this paper due to page limitation, we show the results in Fig. 3 of \cite{Lee:throuOutage} that our analysis can effectively characterize (with small gap) the throughput--outage tradeoff even with finite dimensional setups.

\section{Simulation Results with Practical Setup}

In this section, we present simulation results of the throughput--outage tradeoff considering the practical MZipf popularity distributions and network setup as in \cite{Ji:Dcache} to support the theory. Communications between users occur at 2.45 GHz. For the simulations, we assume a cell of dimensions  $0.36 {\rm km}^2$ ($600{\rm m} \times 600{\rm m}$) that contains buildings as well as streets/outdoor environments. We assume a number of users in the cell $N=10000$, i.e., on average, there are $2 \sim 3$ nodes, every square $10 \times 10$ meters. The cell contains a Manhattan grid of square buildings with side length of $50$m, separated by streets of width $10$m. Each building is made up of offices; of size $6.2 {\rm m} \times 6.2 {\rm m}$. Within the cell, users (devices) are distributed at random according to a uniform distribution. Due to our geometrical setup, each node is assigned to be outdoors or indoors, and in the latter case placed in a particular office. Since $2.4$ GHz communication can penetrate walls, we have to account for different scenarios, which are indoor communication (Winner model A1), outdoor-to-indoor communication (B4), indoor-to-outdoor communication (A2), and outdoor communication (B1) (see \cite{Ji:Dcache}).

The number of clusters in a cell is varied from $2^2, 3^2,....27^2$; a frequency reuse factor of $4$ is used to minimize the inter-cluster interference. The capacity of the cache on each device $S$ is kept as a parameter that we will vary in the simulations below. To provide some real-world connections: storage of an hour-long video in medium video quality (suitable for a cellphone) takes about 300 MByte. Thus, storing 100 files with current cellphones is quite realistic, and given the continuous increase in memory size, even storage of 500 files is not prohibitive (assuming, of course, some incentivization by network operators or other entities). 

In terms of channel models, we mostly employ the Winner channel models with some minor modifications. In particular, we directly use the respective Winner II channel models with antenna heights of $1.5$m, as well as the probabilistic LOS model for Winner. We add a probabilistic body shadowing loss ($\sigma_{L_{b}}$) with a lognormal distribution, where for LOS, $\sigma_{L_{b}} = 4.2$ and for NLOS, $\sigma_{L_{b}} = 3.6$ to account for the blockage of radiation by the person holding the phone; see \cite{karedal2008measurement}.  More details about the channel model can be found in \cite{Ji:Dcache}.
 
\begin{figure}
\centering
\includegraphics[width=7.0cm,keepaspectratio]{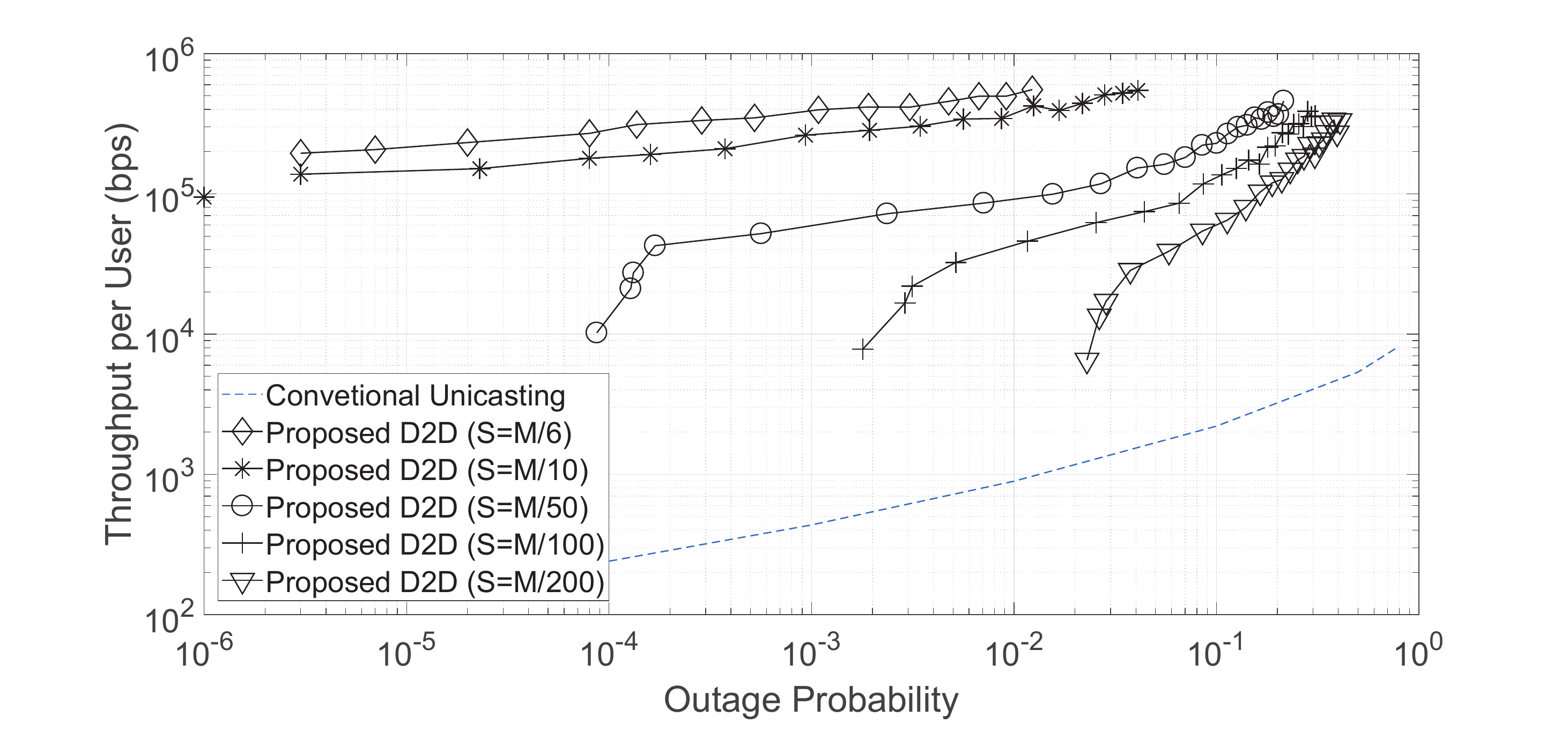}
\caption{Throughput--outage tradeoff in region 2 assuming mixed office scenario; varying local storage size. }
\vspace{-10pt}
\label{fig:throughput_outage_2}
\end{figure}

\begin{figure}
\centering
\includegraphics[width=7.0cm,keepaspectratio]{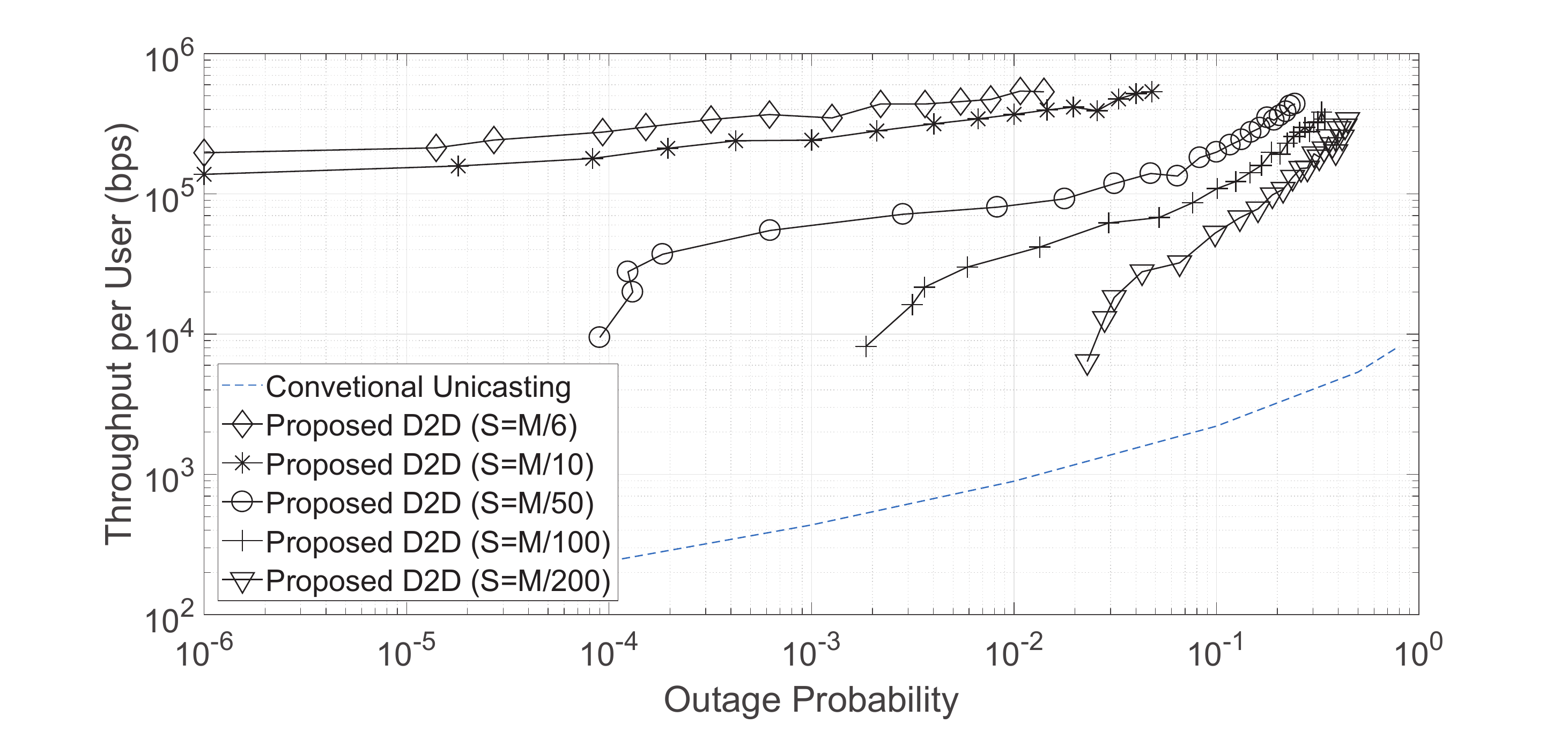}
\caption{Throughput--outage tradeoff in region 3 assuming mixed office scenario; varying local storage size.}
\vspace{-10pt}
\label{fig:throughput_outage_3}
\end{figure}

Figure \ref{fig:throughput_outage_2} shows the throughput-outage tradeoff for different cache sizes on each device in region 2. Outage means here that a file cannot be found in the cache of any device within the cluster of the requesting user, and thus has to be provided via the base station. An outage of $10\%$ thus implies that $90 \%$ of traffic can be offloaded to the D2D connections. We can see that extremely high throughput can be achieved if the cache size of each user is up to $1/10$ of the library size. Even for $S=M/50$, i.e., approx. 100 files (30 GB), the advantage compared to conventional unicasting described in \cite{Ji:Dcache} is two orders of magnitude. Even just the caching of 30 files ($M/200$) provides significant gains, though only for outage probabilities $>0.01$.  The results for region 3 (Fig. \ref{fig:throughput_outage_3}) are very similar.

\section{Conclusions}

In this paper, we have analyzed and evaluated, for the first time, the throughput--outage tradeoff that can be achieved with caching-based D2D communications in real-word scenarios of cellular networks. Our analysis and evaluations adopt the MZipf video popularity distribution that is based on the {\em measured} demand in the biggest video service in the UK, extracted for {\em cellular users only}. The theoretical analysis shows an achievable scaling law no worse than those scaling laws reported in existing literature. We also found that, in simulations adopting a realistic setup, caching-based D2D indeed provides one to two orders of magnitude throughput improvement at outage probabilities between $0.01$ and $0.1$. Therefore, both theoretical and numerical results indicate that the promising benefits of the caching-based D2D networks can be retained for mobile users considering practical popularity distributions.

\bibliographystyle{IEEEtran}
\nobibliography{IEEEabrv}

\end{document}

%% file: macros.tex
\long\def\comment#1{}

\newcommand{\be}{\begin{equation}}
\newcommand{\ee}{\end{equation}}
\newcommand{\bea}{\begin{eqnarray}}
\newcommand{\eea}{\end{eqnarray}}

